\documentclass[amsmath,amssymb,aps,pre,groupedaddress,superscriptaddress,twocolumn]{revtex4}
\usepackage[utf8]{inputenc}
\usepackage[english]{babel}
\usepackage[english]{varioref}
\usepackage{graphicx}
\usepackage{dcolumn}
\usepackage{bm}
\usepackage{amsmath}
\usepackage{amssymb}
\usepackage{caption}
\usepackage[colorlinks=true,linkcolor=blue,urlcolor=blue,citecolor=blue,anchorcolor=blue]{hyperref}
 
\begin{document}

\title{Active and inactive quarantine in epidemic spreading on adaptive activity-driven networks}

\author{Marco Mancastroppa}
\author{Raffaella Burioni} 
\affiliation{Dipartimento di Scienze Matematiche, Fisiche e Informatiche, Università degli Studi di Parma, Parco Area delle Scienze, 7/A 43124 Parma, Italy}
\affiliation{INFN - Istituto Nazionale di Fisica Nucleare, Gruppo Collegato di Parma, Parco Area delle Scienze 7/A, 43124 Parma, Italy}
\author{Vittoria Colizza}
\affiliation{INSERM - Institut national de la santé et de la recherche médicale, Sorbonne Université, Institut Pierre Louis d'Epidémiologie et de Santé Publique (IPLESP), 75012 Paris, France}
\author{Alessandro Vezzani}
\email{alessandro.vezzani@unipr.it}
\affiliation{IMEM-CNR, Parco Area delle Scienze 37/A 43124 Parma, Italy}
\affiliation{Dipartimento di Scienze Matematiche, Fisiche e Informatiche, Università degli Studi di Parma, Parco Area delle Scienze, 7/A 43124 Parma, Italy}

\date{\today}

\begin{abstract}
We consider an epidemic process on adaptive activity-driven temporal networks, with adaptive behaviour modelled as a change in activity and attractiveness due to infection. By using a mean-field approach, we derive an analytical estimate of the epidemic threshold for SIS and SIR epidemic models for a general adaptive strategy, which strongly depends on the correlations between activity and attractiveness in the susceptible and infected states. We focus on strong social distancing, implementing two types of quarantine inspired by recent real case studies: an active quarantine, in which the population compensates the loss of links rewiring the ineffective connections towards non-quarantining nodes, and an inactive quarantine, in which the links with quarantined nodes are not rewired. Both strategies feature the same epidemic threshold but they strongly differ in the dynamics of active phase. We show that the active quarantine is extremely less effective in reducing the impact of the epidemic in the active phase compared to the inactive one, and that in SIR model a late adoption of measures requires inactive quarantine to reach containment.
\end{abstract}

\maketitle

\textit{Introduction.} 
Understanding the effects of changes in the behaviour of individuals during epidemics is essential to improve response strategies and to foster containment \cite{satorras2015epidemic,fraser2004,hollingsworth2011mitigation,anderson2020individual,masuda2013predicting}. This problem is nowadays particularly relevant, due to the extraordinary measures, massively introduced to limit recent disease spreading \cite{who_declaration_public,who_situation_report,quarantine_Italy2,quarantine_China}. In the presence of epidemics, people adapt their behaviour, modifying their actions in several ways. Infected individuals partially or totally reduce their activity, due to the appearance 
of symptoms or, if they are asymptomatic, due to limitations. Analogously, infected individuals can experience a reduction in their attractiveness, due to the self-protective behaviour of other individuals, who try to avoid contacts with the infected ones. All these adaptive behaviours lead to a change in activity, which can be realized in different ways, due to the peculiar structure of the society or to the perception of severity. As an example, as seen in recent real case studies \cite{anderson2020individual,active_Italy}, depending on the possibility to implement strong containment measures, limitation of activity in a population can be implemented in an {\it active} or {\it inactive} way. In the former, individuals avoid contacts with infected nodes, but they can readdress their activity towards non infected individuals, only shifting their focus. In the latter, individuals really become less active, in the sense that if a link is to be established with an infected node, the individual decide not to take the action, therefore reducing the overall activity of the system. A modelling framework of these effects could help in quantifying the relevance of quarantine measures \cite {gross2006adaptive,lagorio2011}, and 
to assess how important is the rapidity of adoption of adaptive measures to reach containment. Simplified models, amenable to analytic control but still including the relevant dynamics, can be of some help here, as they allow testing the relevant parameters best-affecting disease containment. The natural framework to model epidemics is that of temporal networks, where links between nodes evolve in time on the same time scale of the dynamical process \cite{satorras2015epidemic,holme2013temporal,valdano2018epidemic,holme2014birth,perra2012activity,liu2014controlling,ping2018epidemic,rizzo2014effect,moinet2018effect,starnini2013topological,masuda2013predicting}. 

We focus here on the susceptible-infected-susceptible (SIS) and susceptible-infected-recovered (SIR) models on activity-driven networks \cite{perra2012activity,starnini2013topological,ubaldi2016asymptotic,ubaldi2017burstiness,burioni2017asymptotic,tizzani2018memory,mancastroppa2019,kim2018memory} with adaptive behaviour \cite{funk2010review,feniche2011adaptive}. In activity-driven networks, the propensity to engage an interaction is modelled by assigning to each node an activity potential, measuring the typical number of activations (link formations) per time performed by the agent, plus an attractiveness, describing how likely it is for a node to be contacted by others \cite{pozzana2017attractiveness}. The adaptive behaviour results in a change in activity and attractiveness due to infection, expressed through a general distribution function for activity and attractiveness in infected and susceptible nodes. By using an \textit{activity-attractiveness based mean-field} approach, we derive an analytical estimate of the epidemic threshold for SIS and SIR epidemic models, holding for all active and inactive adaptive strategies. We also obtain analytically the epidemic prevalence of the SIS process (endemic state), while for the SIR active phase we perform numerical simulations. Interestingly, the threshold strongly depends on the correlations between the activities and attractivenesses in the susceptible and infected states. In particular, we focus on \textit{active} and \textit{inactive quarantine}, showing that the two containment measures have the same epidemic threshold, but they strongly differ in the dynamic of the active phase. As a result, the active quarantine is extremely less effective in reducing the impact of the epidemics compared to the inactive one, near to the epidemic threshold. We also uncover the strong effects of early adoption of quarantine measures. We show that early adoption can potentially allow to contain the epidemics with an active quarantine, without affecting the activity of the healthy, while a late adoption requires a strong reduction of the overall activity to reach an effective containment.\\

\textit{The model.} We consider the SIS model on adaptive activity-driven networks \cite{perra2012activity,ubaldi2016asymptotic,pozzana2017attractiveness}: each node can be susceptible ($S$) or infected ($I$) and it is characterized by two activity parameters $a_S,a_I$ and two attractiveness parameters $b_S,b_I$ drawn form the joint distribution $\rho(a_S,a_I,b_S,b_I)$. Node activations occur with a Poisson process, with activation rate $a_S$ or $a_I$, according to the node's status. Initially all $N$ nodes are disconnected and when a node activates it creates $m$ links with $m$ randomly selected nodes (hereafter we set $m=1$): the probability to contact a node with attractiveness $b_i$ (with $i=S,I$) is given by $p_{b_i}=b_i/\alpha$, where $\alpha$ is a normalization factor and depends on the adaptive behaviour as we will discuss. All links are deleted after a time step and the procedure is iterated. If a link connects an infected $I$ and a susceptible $S$ node, a contagion occurs with probability $\lambda$: $S+I \xrightarrow[]{\lambda} 2I$, otherwise nothing happens. Infected nodes recover with rate $\mu$, through a Poissonian process: $S \xrightarrow[]{\mu} I$. 
We call $P_{a_S,a_I,b_S,b_I}(t)$ the probability that a node with activities and attractivenesses $(a_S,a_I,b_S,b_I)$ is infected at time $t$. 

The adaptive behaviour can be modelled in two ways, which we will call \textit{active} and \textit{inactive}. In the active case, an active node connects for sure with one of the other nodes. In this case, the normalization factor $\alpha$ is the average attractiveness of the system at time $t$:
$\alpha=\langle b(t)\rangle=\int da_S \, da_I \, db_S \, db_I \, \rho(a_S,a_I,b_S,b_I) [b_S (1-P_{a_S,a_I,b_S,b_I}(t)) + b_I P_{a_S,a_I,b_S,b_I}(t)]=
 \overline{b_S} + \overline{b_IP}(t)-\overline{b_S P}(t) $ 
 where we define $\overline{f}(t)=\int da_S \, da_I \, db_S \, db_I \, \rho(a_S,a_I,b_S,b_I) f_{a_S,a_I,b_S,b_I}(t)$.  $P_{a_S,a_I,b_S,b_I}(t)$ evolves in time according to the following equation:
\begin{widetext}
\begin{equation}
\partial_t P_{a_S,a_I,b_S,b_I}(t) = -\mu P_{a_S,a_I,b_S,b_I}(t)+\lambda (1-P_{a_S,a_I,b_S,b_I}(t))\frac{a_S \overline{b_IP}(t)+b_S \overline{a_I P}(t)}{\overline{b_S} + \overline{b_IP}(t)-\overline{b_S P}(t)}.
\label{eq:EQ_active}
\end{equation}
 
In the inactive case, an active node may not connect to any of the nodes due to the reduction of the average attractiveness of the system caused by the infection process. In this case, at each time $t$ we set $\alpha=\overline{b_S}$ i.e. the average attractiveness when all sites are susceptible. The probability for an active node not to connect is:
$(\overline{b_S}-\langle b(t)\rangle)/\overline{b_S}$ and the evolution of $P_{a_S,a_I,b_S,b_I}(t)$ is given by: 

\begin{equation}
\partial_t P_{a_S,a_I,b_S,b_I}(t) = -\mu P_{a_S,a_I,b_S,b_I}(t)+\lambda (1-P_{a_S,a_I,b_S,b_I}(t))\frac{a_S \overline{b_IP}(t)+b_S \overline{a_I P}(t)}{\overline{b_S}}.
\label{eq:EQ_inactive}
\end{equation}

Eq.s (\ref{eq:EQ_active},\ref{eq:EQ_inactive}) are exact due to the mean-field nature of the model, since local correlations are destroyed at each time step and since we do not consider memory \cite{tizzani2018memory,kim2018memory} or burstiness effects \cite{mancastroppa2019,ubaldi2017burstiness,burioni2017asymptotic}.

The SIS model features a phase transition between an absorbing and an active phase: the control parameter is the effective infection rate $r = \lambda/\mu$ (in literature the threshold is sometimes discussed in terms of the adimensional parameter $r 2 \langle a \rangle = r 2 \overline{a_S}$). Through a linear stability analysis, we obtain the epidemic threshold $r_C$, which remarkably is the same for the active and inactive adaptive behaviour \cite{supplemental}:
\begin{equation}
r_C=\left. \frac{\lambda}{\mu}\right|_C=\frac{2 \overline{b_S}}{
\overline{a_S b_I} + \overline{a_I b_S} + 
\sqrt {(\overline{a_S b_I} - \overline{a_I b_S})^2 + 4 
\overline{a_S a_I} \, \overline{b_S b_I}}}.
\label{eq:EQ_thr}
\end{equation}
\end{widetext}
From Eq. (\ref{eq:EQ_thr}), we see that $r_C$ is reduced in the presence of strong correlations between susceptible and infected states. In particular, this means that epidemic spreading is favoured by the presence of nodes which keep their high activity and/or high attractiveness both when they are susceptible and when they are infected (super-spreaders).
Eq. (\ref{eq:EQ_thr}) also reproduces the non-adaptive case without attractiveness \cite{perra2012activity} (fixing $a_I=a_S$ and $b_I=b_S=1$) $r_C^{NAd}=( \overline{a_S} + \sqrt{\overline{a_S^2}})^{-1}$ and the non-adaptive case with attractiveness (NAdwA) \cite{pozzana2017attractiveness} (fixing $b_I=b_S$ and $a_I=a_S$) $r_C^{NAdwA}=\overline{b_S} (\overline{a_S b_S} + \sqrt{\overline{a_S^2} \cdot \overline{b_S^2}})^{-1}$. If $b_S=c a_S$ ($c$ arbitrary constant), as observed in real systems \cite{pozzana2017attractiveness}, we obtain $r_C^{NAdwA}=\overline{a_S}/[2 \overline{a_S^2}]$.

From Eq.s (\ref{eq:EQ_active},\ref{eq:EQ_inactive}), the stationary infection probability $P_{a_S,a_I,b_S,b_I}^0=\lim\limits_{t \to \infty} P_{a_S,a_I,b_S,b_I}(t)$ in the active (Eq. \eqref{eq:P_active}) and inactive (Eq. \eqref{eq:P_inactive}) cases is:
\begin{equation}
P_{a_S,a_I,b_S,b_I}^0=\frac{a_S \overline{b_IP}+b_S\overline{a_IP}}{\frac{\mu}{\lambda}(\overline{b_S} + \overline{b_I P} - \overline{b_S P})+a_S \overline{b_IP}+b_S\overline{a_IP}},
\label{eq:P_active}
\end{equation}
\begin{equation}
P_{a_S,a_I,b_S,b_I}^0=\frac{a_S \overline{b_IP}+b_S\overline{a_IP}}{\frac{\mu}{\lambda} \overline{b_S} +a_S \overline{b_IP}+b_S\overline{a_IP}}.
\label{eq:P_inactive}
\end{equation}

Contrarily to the epidemic threshold, the active phase depends on the implementation of the adaptive behaviour.
This suggests that it is not enough to investigate the epidemic threshold, or the basic reproductive number $R_0$: the same population with two different dynamics feature a different epidemic active phase. 

The proposed model allows studying several adaptive behaviours; in particular by setting the functional form of $\rho(a_S,a_I,b_S,b_I)$ one can model a mild social distancing, like the sick-leave practice where only activity is reduced in infected nodes \cite{ariza2018healthcareseeking}; a targeted adaptive prescription where $a_I$ and $b_I$ vary only for some activity classes, more exposed to risk \cite{anderson2020individual,mossong2008social}; moreover one can consider different distributions of $a_S$ and $b_S$ modelling different social systems \cite{pozzana2017attractiveness,perra2012activity}. Here we will focus on a strong social distancing approach like \textit{quarantine}. \\

\textit{Active and inactive quarantine.} Nowadays the effects of quarantine on epidemic spreading are of enormous interest due to the extraordinary isolation measures to limit the spread of COVID-19 taken by several countries \cite{who_declaration_public,who_situation_report,quarantine_China,quarantine_Italy2}.
In our model we consider that a fraction $\delta$ of the nodes, when infected, goes to quarantine by setting both $a_I$ and $b_I$ to zero. 
Indeed, we expect that a fraction of the infected nodes does not perform quarantine since they may not be aware of the infection (no symptoms) or they do not follow the prescriptions of social distancing: thus $\delta$ takes into account the uncertainty in the application of the containment measures. 

Our approach naturally distinguishes an active quarantine from an inactive one. In the active quarantine an active node selects randomly its contact among the non-quarantining nodes: links directed to quarantining nodes are effectively rewired towards potentially susceptible or non-quarantining infected nodes. 
Dynamical link rewiring can produce non-trivial effects on epidemic spreading, as observed in \cite{gross2006adaptive} for static networks.
{ Indeed, rewiring induces new connections among not-quarantining nodes and this can strengthen the epidemics \cite{scarpino2016effect,gross2006adaptive}.
On the contrary, in the inactive case the population does not compensate the ineffective links towards nodes in quarantine and the effective contagion is reduced.}

We consider a system with a linear correlation $b_S = c a_S$, accordingly to observations on real data \cite{pozzana2017attractiveness}. Very active hubs, generating many links, are also very attractive, receiving just as many. If a fraction $\delta$ of the population performs quarantine ($a_I=b_I=0$), while the remaining $1-\delta$ keeps $a_I=a_S$ and $b_I=b_S$, we get $\rho(a_S,a_I,b_S,b_I)= \rho_S(a_S) \delta(b_S-c a_S)[(1-\delta) \delta(a_I-a_S) \delta(b_I-b_S)+ \delta \cdot \delta(a_I) \delta(b_I)]$, where $\delta(x)$ is the Dirac-delta function. The epidemic threshold is:
\begin{equation}
r_C^{quarantine}=\frac{1}{1-\delta}\frac{\overline{a_S}}{2 \overline{a_S^2}}=\frac{r_C^{NAdwA}}{1-\delta}.
\label{eq:soglia}
\end{equation}
The quarantine increases the epidemic threshold of the non adaptive case by a factor $(1-\delta)^{-1}$ which is particularly significant when $\delta \sim 1$. 

Although the active and inactive quarantine display the same epidemic threshold, the active phase is different. The stationary infection probability in the active quarantine is:
\begin{equation}
P_{a_S,a_I,b_S,b_I}^0=\frac{2 a_S \overline{a_SP}}{\frac{\mu}{\lambda} \left( \frac{\overline{a_S}-\overline{a_SP}}{1-\delta}+\overline{a_SP} \right) + 2 a_S \overline{a_S P}},
\label{eq:P_SIS_active}
\end{equation}
while, in the inactive case it is:
\begin{equation}
P_{a_S,a_I,b_S,b_I}^0=\frac{2 a_S \overline{a_SP}}{\frac{\mu}{\lambda}\frac{\overline{a_S}}{1-\delta}+ 2 a_S \overline{a_SP}}.
\label{eq:P_SIS_inactive}
\end{equation}

\begin{figure}
\centering
\includegraphics[width=0.45\textwidth]{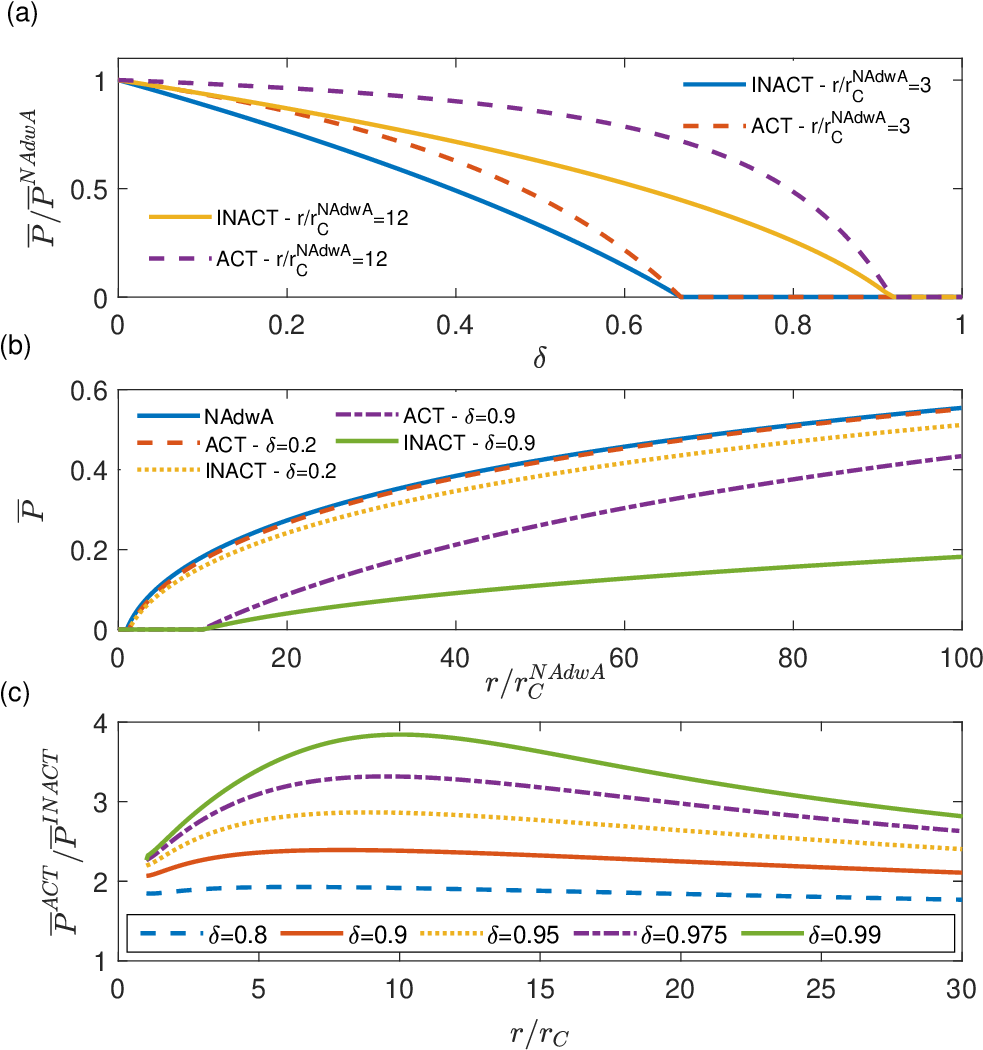}
\caption{Effects of quarantine on SIS active phase. In panel (a) we plot the ratio $\overline{P}/\overline{P}^{\rm NAdwA}$ as a function of $\delta$ for the active and the inactive quarantine fixing $r=3\cdot r_C^{\rm NAdwA}$ and $r=12\cdot r_C^{\rm NAdwA}$. In panel (b) we plot the prevalence $\overline{P}$ as a function of $r/r_C^{\rm NAdwA}$, for the NAdwA case and for the active/inactive quarantines, fixing $\delta=0.2$ and $\delta=0.9$. In panel (c) we plot the ratio $\overline{P}^{ACT}/\overline{P}^{INACT}$ between the epidemic prevalence in the active and inactive case as a function of $r/r_C$ for several values of $\delta$ (with $r_C$ the threshold for the adaptive case). In all panels, $\nu=0.5$, $a_m=10^{-3}$, $a_M=1$.}
\label{fig:SIS_prev}
\end{figure}

Eq.s (\ref{eq:P_SIS_active},\ref{eq:P_SIS_inactive}) can be solved self consistently fixing $\rho_S(a_S)$: we consider a power-law distribution $\rho_S(a_S) \sim a_S^{-(\nu+1)}$ with $a_S \in [a_m,a_M]$ \cite{perra2012activity,ubaldi2016asymptotic}, modelling the presence of heterogeneities and large activity fluctuations. Indeed, many real human systems feature a broad power-law distribution of $a_S$ with exponent $\nu \sim 0.3-1.5$ \cite{ubaldi2016asymptotic,perra2012activity, karsai2014time}. Heterogeneities account for different social propensity in engaging social interactions (e.g. different works, social roles), producing different number of contacts over time \cite{perra2012activity,ubaldi2016asymptotic}: in the following we will fix $\nu=0.5$.

In Fig.~\ref{fig:SIS_prev}(a) we plot the ratio between epidemic prevalence $\overline{P}$ and the prevalence in the NAdwA case (i.e. at $\delta=0$) as a function of $\delta$; quarantine measures induce an effective reduction of the final prevalence (of a factor $\gtrsim 2$) only for $\delta \gtrsim 0.5$.
This is confirmed by Fig.~\ref{fig:SIS_prev}(b) where we plot the prevalence as a function of $r/r_C^{NAdwA}$ for $\delta=0.2$ and $\delta=0.9$, so here we set $\delta$ in the range $0.7-1$, consistently with the recent implementations of extended quarantine \cite{quarantine_China,quarantine_Italy2}.
Interestingly, Figures \ref{fig:SIS_prev}(a-b) show that the epidemic prevalence of the inactive quarantine is much smaller compared to the active one, making the latter strategy extremely less effective.
In this respect, in Fig. \ref{fig:SIS_prev}(c) we plot the ratio between the epidemic prevalence of the active and inactive case as a function of $r/r_C$. The difference is maximized when $r/r_C \sim 5-10$ (depending on $\delta$) and it increases with higher $\delta$ values. Also for more realistic $r/r_C \sim 1$ values, where the quarantine is effective in moving the systems near to the critical point, in active quarantine the epidemic prevalence is about twice that of the inactive one.\\

\textit{Effects of quarantine on SIR epidemic model.}
In the SIR model, a recovered node is no more susceptible but enters at the recovery rate $\mu$ into the immune state $R$ where it cannot be infected: $I \xrightarrow[]{\mu} R$. We consider the SIR model on our activity-driven model: each node is described by six parameters $(a_S,a_I,a_R,b_S,b_I,b_R)$ extracted from the joint distribution $\rho_{SIR}(a_S,a_I,a_R,b_S,b_I,b_R)$.
Hereafter, we consider the case in which all recovered nodes behave as if they were susceptible $a_R=a_S$ and $b_R=b_S$: $\rho_{SIR}(a_S,a_I,a_R,b_S,b_I,b_R)=\rho(a_S,a_I,b_S,b_I) \delta(b_R-b_S) \delta(a_R-a_S)$, with $\rho(a_S,a_I,b_S,b_I)$ equal to the previous distribution for quarantine.
 
Due to the mean-field nature of the model, the epidemic threshold is the same of the SIS model independently of $a_R$ and $b_R$ and it is equal for active and inactive quarantine (Eq. \ref{eq:soglia}). On the contrary, the dynamics in the active phase is known to be different. In particular,
the SIR model lacks an endemic steady state, since the dynamics always halts reaching a state in which no infected nodes are present. The final fraction of recovered nodes $R_\infty$ is used as an order parameter. However the stationary condition on the dynamical equation does not provide a solution for $R_\infty$ and numerical simulation are necessary to obtain insight on the dynamics. 

We implement the SIR model with a continuous time Gillespie-like algorithm \cite{gillespie1976general,supplemental}, in a system of $N$ nodes and we average over a number of realizations of the dynamical evolution and of the disorder, so that the error on $R_{\infty}$ is lower than 1\%. The initial conditions are imposed infecting the node with the highest activity $a_I$ \cite{boguna2013}, immediately implementing the quarantine measures.

\begin{figure}
\centering
\includegraphics[width=0.45\textwidth]{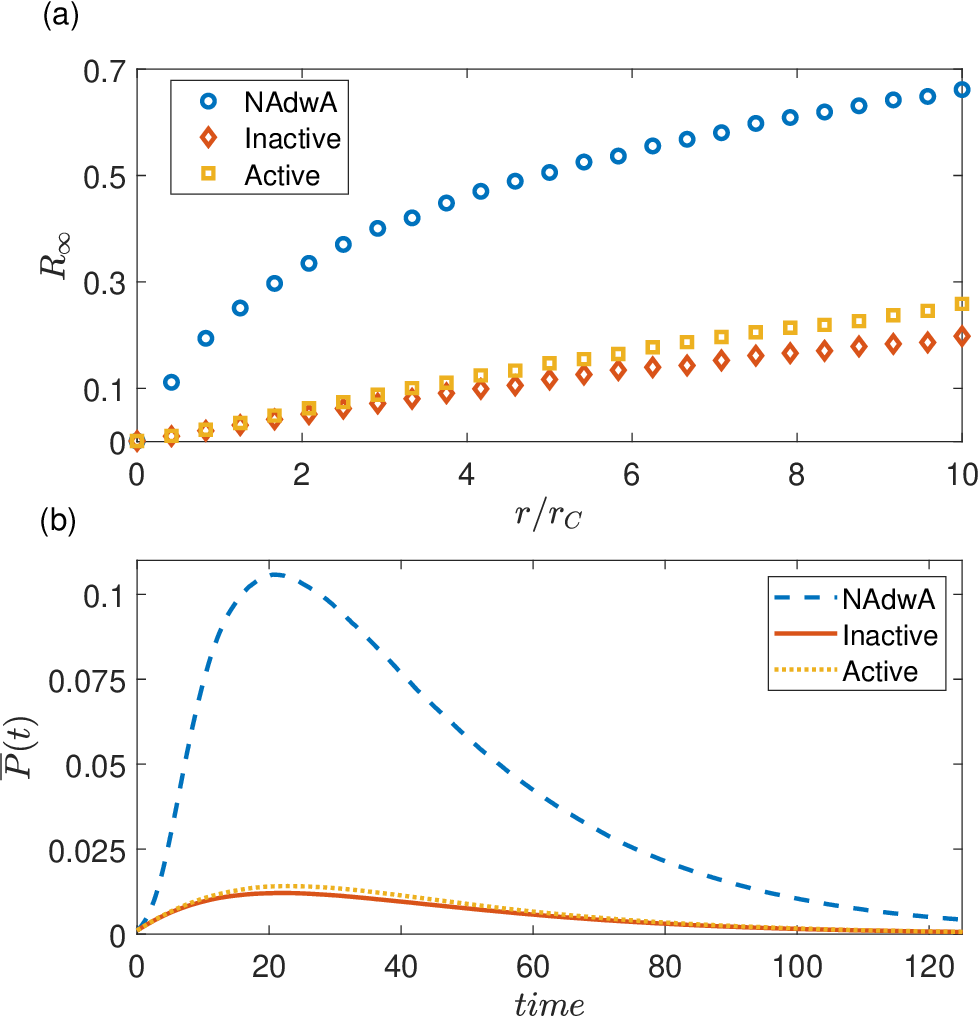}
\caption{Effects of quarantine on the SIR active phase. In panel (a) we plot the epidemic final-size $R_{\infty}$ as a function of $r/r_C$ (with $r_C$ the threshold for the adaptive case), for the NAdwA case and for the active/inactive quarantines. In panel (b) we plot the temporal evolution of the average infection probability $\overline{P}(t)$ for the NAdwA case and for the active/inactive quarantines, fixing $r/r_C=1.4$. In both panels, $\delta=0.9$, $N=10^3$, $\nu=0.5$, $a_m=10^{-3}$, $a_M=1$}
\label{fig:SIR_prev}
\end{figure}

In Fig.~\ref{fig:SIR_prev}(a) we compare $R_{\infty}$ of the NAdwA case with the active/inactive quarantines, computed as a function of $r/r_C$, fixing the same $\delta=0.9$. Both quarantines deeply lowers the epidemic final-size compared to the non-adaptive case, however the active quarantine produces an higher $R_{\infty}$ than the inactive one. The epidemic final-size of the active case is about 10\% higher than the inactive one for small infectivity $r/r_C \sim 1$, i.e. the significant regime for an effective quarantine. Both quarantine strategies greatly impact on the SIR dynamics: Fig.~\ref{fig:SIR_prev}(b) shows the temporal evolution of the average infection probability $\overline{P}(t)$ for $\delta=0.9$ and $r/r_C = 1.4$. { The infection peak is strongly flattened and occurs slightly before, compared to the NAdwA dynamics. This is expected, as social distancing anticipates the decay of the fraction of infected nodes. 
The height of the peak for the active quarantine is 10\%-20\% larger than the inactive one for $r/r_C \sim 1$ (depending on $\delta$ value).} These estimates outlines a difference between the two strategies even if the effect is smaller compared to the SIS model.

Thus far we considered quarantine measures implemented at the beginning of the epidemic. However typically the containment measures are applied only after a fraction $\beta$ of the population has been infected. This is relevant for the SIR model, since the dynamics and $R_\infty$ depend strongly on the initial conditions, contrary to the SIS.
Therefore, we let the epidemic process evolve on the NAdwA network, without adaptive behaviours, until a fraction $\beta$ of the population has been infected. Then, quarantine measures are implemented and are kept effective until the end of the epidemic.

\begin{figure}
\centering
\includegraphics[width=0.45\textwidth]{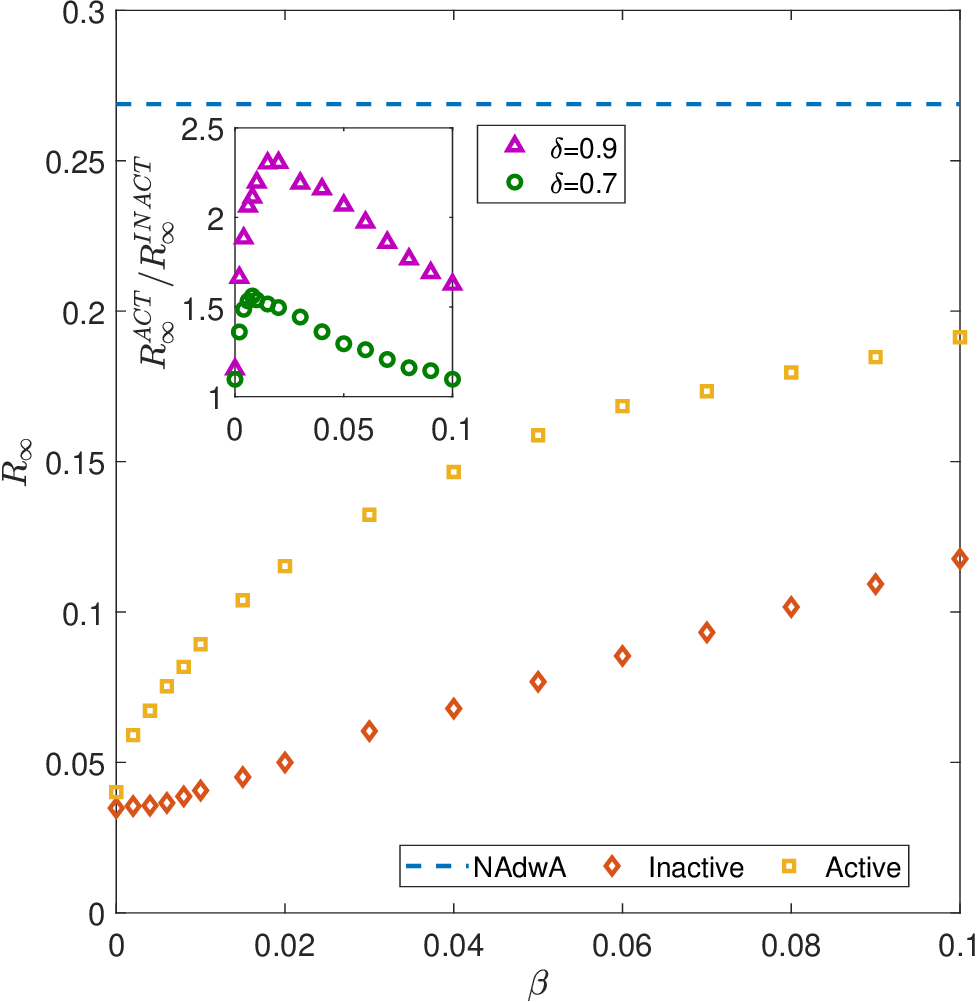}
\caption{Effects of timing in quarantine implementations. We plot the epidemic final-size $R_{\infty}$ as a function of $\beta$ fraction of nodes been infected before quarantine measures are implemented. We consider active and inactive quarantines, fixing $\delta=0.9$ and we compare them with the NAdwA case. In the insert we plot the ratio $R_{\infty}^{ACT}/R_{\infty}^{INACT}$ between the epidemic final-size in the active and inactive case as a function of $\beta$ for several $\delta$ values. In both panels, $r/r_C=1.4$ (with $r_C$ the threshold for the adaptive case), $N=10^3$, $\nu=0.5$, $a_m=10^{-3}$, $a_M=1$.}
\label{fig:SIR_fraz_Rinf}
\end{figure}

\begin{figure}
\centering
\includegraphics[width=0.45\textwidth]{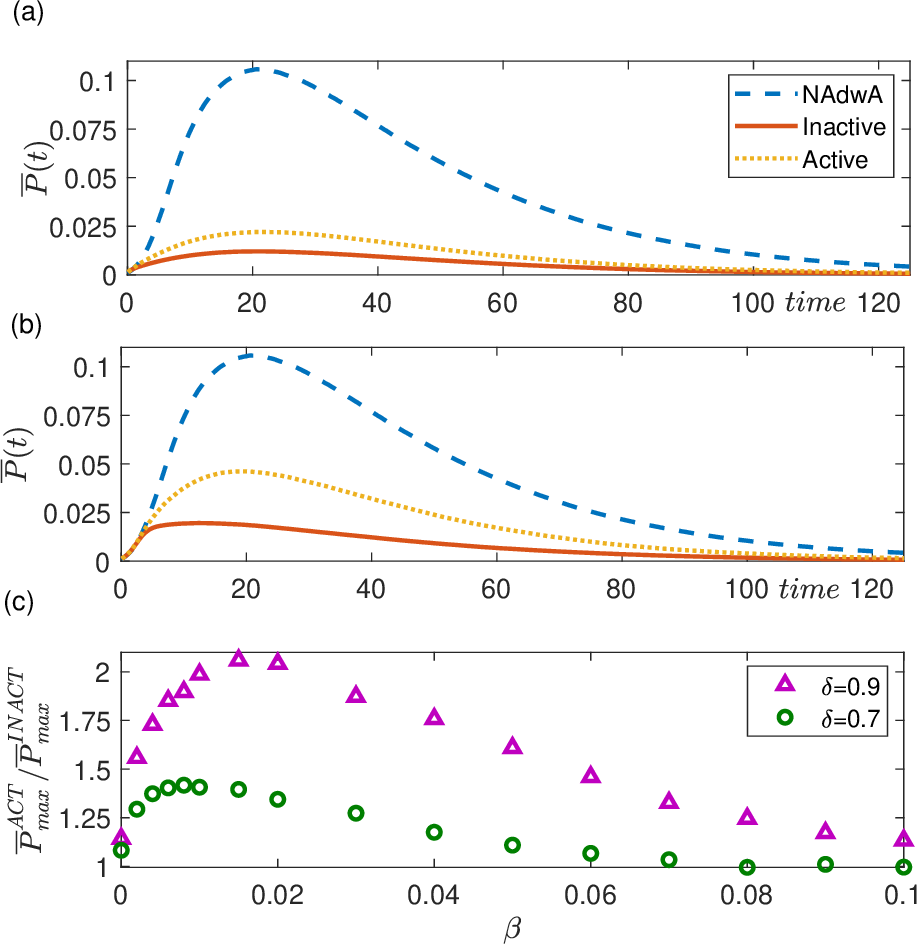}
\caption{Effects of timing in quarantine implementations. In panel (a) and (b) we plot the temporal evolution of the average infection probability $\overline{P}(t)$ for the NAdwA case and for active/inactive quarantine, fixing $\delta=0.9$ and respectively $\beta=0.0025$ and $\beta=0.02$. In panel (c) we plot the ratio $\overline{P}_{max}^{ACT}/\overline{P}_{max}^{INACT}$ between the height of the infection peak in the active and inactive case as a function of $\beta$, for several values of $\delta$. In all panels, $r/r_C=1.4$ (with $r_C$ the threshold for the adaptive case), $N=10^3$, $\nu=0.5$, $a_m=10^{-3}$, $a_M=1$.}
\label{fig:SIR_fraz_picco}
\end{figure}

In Fig.~\ref{fig:SIR_fraz_Rinf} we plot $R_{\infty}$ as a function of $\beta$, fixing $\delta=0.9$ and $r/r_C=1.4$: both strategies show the importance of an early adoption of quarantine, since $R_{\infty}$ increases significantly with $\beta$, despite in a different way for the two strategies. In the inactive quarantine approximatively we get $R_{\infty}(\beta) \sim R_{\infty}(\beta=0)+\beta$, i.e. the epidemic final-size for the quarantine at $\beta = 0$ is summed to the fraction $\beta$ of infected nodes before the quarantine adoption. On the contrary, in active quarantine, $R_{\infty}$ grows very rapidly with $\beta$ and the difference between the two strategies in weakening the epidemic becomes very different. 
In the insert of Fig.~\ref{fig:SIR_fraz_Rinf}, we show the fast growth with $\beta$ of the ratio between the epidemic final-size of the active and inactive case. 
If the quarantine is implemented immediately ($\beta=0$), the difference is about 10\%; if the containment measures are applied when about 2\% of the population has been infected $R_{\infty}$ for the active case can be up to $2.5$ times the inactive one (for $\delta =0.9$).

Finally, we investigate the effects of the timing in the quarantine adoption on the SIR temporal dynamics: in Fig.~\ref{fig:SIR_fraz_picco}(a)-(b) we plot the infection peak respectively for $\beta=0.0025$ and $\beta=0.02$, fixing $\delta=0.9$ and $r/r_C=1.4$. If the fraction $\beta$ increases, both quarantines are extremely less effective in flattening the infection peak, showing the importance of early adoption of measures. Comparing these results with Fig.~\ref{fig:SIR_prev}(b) ($\beta=0$), for $\beta>0$ the inactive infection peak is significantly flattened compared to the active one. This is highlighted in Fig.~\ref{fig:SIR_fraz_picco}(c) where we plot the ratio between the height of the infection peak in the active and inactive case. If the containment measures are applied immediately, the difference is about 10\%; if they are applied when a small fraction between 0.25\% and 4\% of the population has been infected, the active peak is about twice larger than the inactive one at $\delta = 0.9$; a smaller but significant difference is observed also for $\delta =0.7$.\\

Our results hold for a realistically heterogeneous population \cite{perra2012activity} with extensive containment measures, such as those recently implemented \cite{quarantine_China,quarantine_Italy2}, able to move the system just above the threshold ($r/r_C=1.4$). In particular, in this range of parameters, our findings support the crucial role of timing in the adoption of containment measures. If the measures are implemented immediately, both active and inactive strategies are effective in reducing the infection peak of 75\%-85\% (depending on $\delta$) compared to the NAdwA case. However, any delay in the adoption of quarantine measures produces a drastic reduction in the effectiveness of the strategies. More interestingly, the differences between the two strategies increases in the case of a late adoption. { This means that if containment measures are not immediately taken, a stringent inactive quarantine, also binding the healthy to decrease their activity, is required to produce an effective result. With an early adoption, of course an inactive quarantine is still the best choice, but if this strong social distancing cannot be implemented for social reasons, a strategy focusing only on infected individuals is also able to produce a containment. Such effective results in both cases depend on the ability to obtain a large adhesion to quarantine measures ($\delta=0.7-0.9$).}

\end{document}